\documentclass[11pt, epsf]{article}
\usepackage{epsfig}
\def\ga{\mathrel{\raise.3ex\hbox{$>$\kern-.75em\lower1ex\hbox{$\sim$}}}}
\def\la{\mathrel{\raise.3ex\hbox{$<$\kern-.75em\lower1ex\hbox{$\sim$}}}}

\def\I_M{{I_{\scriptscriptstyle M\times M}}}

\begin{document}

\thispagestyle{empty}
\rightline{IP/BBSR/2004-15}

\vskip 2cm \centerline{ \Large \bf Constructing more non-singular 
$Sp$-branes}

\vskip .2cm

\vskip 1.2cm

\centerline{ \bf Anindya Biswas\footnote{E-mail: 
anindyab@iopb.res.in}}
\vskip 10mm \centerline{ \it Institute of Physics, 
Bhubaneswar-751 005, India} 
\vskip 1.2cm
\vskip 1.2cm
\centerline{\bf Abstract}
\noindent
In this note, we construct an array of {\it non-singular} $Sp$ (where
$p=D-4$) branes in arbitrary $D$ dimensions starting from static 
solutions of black $p$ brane. These solutions 
carry nontrivial time dependent profiles of
dilaton and associated form fields. We also
study other non-singular time dependent configurations. These are  
obtained from proper analytic continuations of {\it non-extremal} diholes.
\newpage
\setcounter{footnote}{0}
\noindent
Non-singular time dependent solutions are very rare in gravity
theories, be it supergravity or string theory. Therefore, the recent
constructions of non-singular $Sp$-brane solutions \cite{strominger}
have attracted much attention \cite{wang}. $Sp$ branes are time dependent
gravitational configurations with $p$ longitudinal directions
\cite{Kru}. However, most of these solutions are singular\footnote
{Construction of $S$-branes through double Wick rotations are
discussed in \cite{Su}, but all these solutions suffer from
singularities. For earlier construction of $S$ brane solutions, 
see \cite{Roy}}. In a recent paper
\cite{strominger}, non-singular $S0$ brane solutions are constructed by 
applying double-Wick rotations \cite{strominger} to known black diholes
\cite{Emparan2}. Analytic continuation of these solutions give
exact, time dependent, non-singular $S0$-brane in four
dimensions. These solutions are non-singular in the sense that, all
the conical singularities, which were there in the static solutions of
black diholes, can be pushed off to the complex plane of
$\rho$ and $t$\footnote{$\rho$ is the radial direction of cylindrical 
polar coordinate system and $t$ is the time coordinate.}.
These solutions often lead to accelerating and
bouncing cosmologies. In this note, we generalise these non-singular 
solutions by analytically continuning $p$ branes of
\cite{LuPope}. We construct $(D-4)$ dimensional $Sp$-brane solutions
in $D$ dimensions. These are the solutions of $D$ dimensional maximal
supergravity theories and hence can be embedded in string theories. We
also construct other non-singular time dependent configurations by 
analytically continuing the non-extremal diholes of 
\cite{Manko}, \cite{Emparan1}. \\\\
\noindent
This letter is organised as follows. First, we construct the non-singular 
$Sp$ branes by Wick rotating the solutions of
\cite{LuPope}. Here, we give various plots to expose the non-singular
nature of the solutions. Subsequently, we analyse the double analytic
continuation of {\it non-extremal} generalisation of black dihole
which are different from \cite{strominger}, following the 
solutions given in \cite{Manko},
\cite{Emparan1}. This allow us to construct non-singular $S0$ branes.\\\\
\noindent
A general class of non-extremal black $p$ 
brane solutions in arbitrary $D$ dimensional space-time was
constructed in \cite{LuPope}. 
The center of the brane sits along the 
transverse $z$ axis. We take the linear superposition (unlike the
non-linear superposition in the case of non-extremal black
dihole) of the $N$ individual $p$-brane potentials to get the potential 
of $N$ number of branes along the $z$ axis. As non-extremal branes
exert forces on each other, the superposition of these branes is a
tricky issue. This can be achieved however, if many of them are
organised on a circle. This has been discussed in \cite{LuPope}.\\\\
\noindent
We start with the general supergravity action in $D$ dimensions of the form
\begin{equation}
S=\int{d^D x \sqrt{-g} \Big(R-{1\over
2}(\partial\phi)^2-{1\over {2(p+2)!}}e^{-a\phi}{F^2}_{(p+2)}\Big)}.
\end{equation}
Here $\phi$ is the dilaton field, and $F_{(p+2)}$ is the antisymmetric 
$(p+2)$ form field strength. The constant $a$ denotes the dilaton 
coupling to the form field $F_{(p+2)}$. From this action we can construct
a static solution which corresponds to $N$ $(D-4)$ branes organised along
a common transverse axis. Below, we present the solution; for more detail
we refer the reader to the original paper \cite{LuPope}. \\
\begin{equation}
ds_D^2 = e^{{4\over{D-2}}(U-\tilde U)}(- e^{2\tilde U} d\tau^2 +
dx_idx^i) + e^{2({D-4\over{D-2}}){\tilde U}
-{4({D-3\over{D-2}})U}}[e^{2K}(d\rho^2 + dz^2)+ \rho^2 d\theta^2],
\label{metric3}
\end{equation}
\begin{equation}
e^{2\tilde U} = {\prod_{n=1}^N} {{r_n+\tilde r_n - k_n}\over {r_n+
    \tilde r_n +k_n}},
\label{e2u}
\end{equation}
\newpage
\begin{eqnarray}
e^{2K}= {\prod_{n=1}^{N}}{\prod_{m=n+1}^{N}} &\Big[&{{r_m\tilde r_n 
+ (z-z_m - {1\over 2}k_m)(z-z_n + {1\over 2}k_n)+\rho^2}\over{r_m r_n
+ (z-z_m-{1\over2}k_m)(z-z_n - {1\over 2}k_n)+\rho^2}}\Big] \times \cr
& \cr
&\Big[&{{r_n\tilde r_m + (z-z_m + {1\over 2}k_m)
(z-z_n - {1\over 2}k_n)+\rho^2}\over{\tilde
r_n\tilde r_m + (z-z_m + {1\over 2}k_m)
(z-z_n + {1\over 2}k_n)+\rho^2}}\Big],
\label{e4k}
\end{eqnarray}
\begin{eqnarray}
r_n = {\sqrt{\rho^2+(z-z_n - {1\over 2}k_n)^2}},~~~~~~~~~~~
{\tilde r_n} = {\sqrt{\rho^2+(z-z_n + {1\over 2}k_n)^2}},
\label{R3}
\end{eqnarray}
\begin{eqnarray}
e^{-\Delta U} &=& (e^{-\tilde U} - c^2 e^{\tilde U})e^{-a^2 \tilde
U},~~~~~~~~~~{\phi}_D = 2 a_D (U-\tilde U),
\label{u,Phi}
\end{eqnarray}
\begin{equation}
A_{0i_1...i_p} = 2ce^{2 \tilde U}(1-c^2 e^{2\tilde U})^{-1}
\epsilon_{i_1...i_p}.
\label{u,A,Phi}
\end{equation}
\noindent
In (\ref{metric3}) $\tau$, $x^i$ are the $p+1$ world volume coordinates
and $\rho, z, \theta$ are the transverse coordinates.
Branes are of masses $k_n=2 M_n$ (for $n$th brane) and centered at points 
$z_n=nb$ (that is several branes are arranged periodically along 
the symmetry axis as discussed earlier).
$r_n$ and $\tilde r_n$ are the distances from the brane center to the 
field point. 
Here we would like to make some comments on $\Delta$, where
$\Delta = a^2 + {2(p+1)(D-p-3)\over(D-2)}$ according to Ref.\cite{Lu}.
If we consider black hole ($p=0$) solution in pure Einstein-Maxwell 
theory where there is no dilaton
coupling in the action, then $a=0$ and $\Delta=1$\footnote{for
more detail see \cite{Pope}}. We get back
the array of standard neutral black hole solution by
putting $c=0$ (in general $c$ can take values other than zero). 
Analytic continuation then gives $S0$-branes of 
\cite{strominger}. For different values of $a$, we get different
supergravity theories. In this note we will focus on the solutions with
single scalar $\phi_D$, but there can be other solutions with
more than one non-zero scalar field excited. \\\\
\noindent
The double analytic continuation, $\tau\rightarrow{iy}$ 
and $z\rightarrow{it}$, changes the solutions to a time 
dependent one. \\
\begin{equation}
ds_D^2 = e^{{4\over{D-2}}(U-\tilde U)}(e^{2\tilde U} dy^2 +
dx_idx^i) + e^{2({D-4\over{D-2}}){\tilde U}
-{4({D-3\over{D-2}})U}}[e^{2K}(d\rho^2 - dt^2)+ \rho^2 d\theta^2],
\label{metric4}
\end{equation}
\begin{equation}
e^{2\tilde U} = {\prod_{n=1}^N} {{r_n+\tilde r_n - k_n}\over {r_n+
\tilde r_n +k_n}},
\label{eu2}
\end{equation}
\begin{eqnarray}
e^{2K}= {\prod_{n=1}^{N}}{\prod_{m=n+1}^{N}} &\Big[&{{r_m\tilde r_n 
- (t+i(z_m + {1\over 2}k_m))(t+i(z_n - {1\over 2}k_n))+\rho^2}\over{r_m r_n
- (t+i(z_m+{1\over2}k_m))(t+i(z_n + {1\over
2}k_n))+\rho^2}}\Big] \times  \cr
& \cr
&\Big[&{{r_n\tilde r_m - (t+i(z_m - {1\over 2}k_m))
(t+i(z_n + {1\over 2}k_n))+\rho^2}\over{\tilde
r_n\tilde r_m - (t+i(z_m - {1\over 2}k_m))
(t+i(z_n - {1\over 2}k_n))+\rho^2}}\Big],
\label{e42}
\end{eqnarray}
\begin{eqnarray}
r_n = \sqrt{\rho^2 - (t + i(z_n + {1\over 2}k_n))^2}, ~~~~~~~
\tilde r_n = \sqrt{\rho^2 - (t + i(z_n - {1\over 2}k_n))^2}.
\label{R4}
\end{eqnarray}\\
\newpage
\noindent
This is a non-singular $Sp$ $(p=D-4)$ brane, as can be seen 
from the figures below. Here, we plot various metric components 
for $N=2$ and $\Delta=4$ (for $M_n<b$). This value of $\Delta$
appears in the ten dimensional type $IIA$ supergravity theory where
the dilaton coupling $a$ to the eight form field strength is 
${3\over 2}$; see for example \cite{Duff}. 
\begin{figure}[ht]
\epsfxsize=8cm
\centerline{\epsfbox{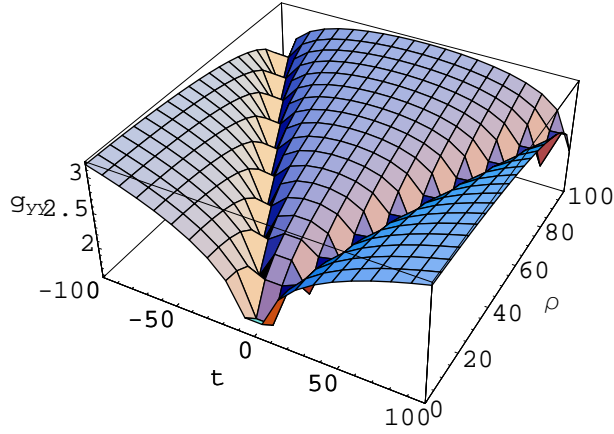}}
\caption{Plot of $g_{yy}$ vs. $\rho$ and $t$ for $D=10$, $p=6$,
$b=1$, $c=1$ and $M_n={1\over 2}$}
\end{figure} 
\begin{figure}[ht]
\epsfxsize=8cm
\centerline{\epsfbox{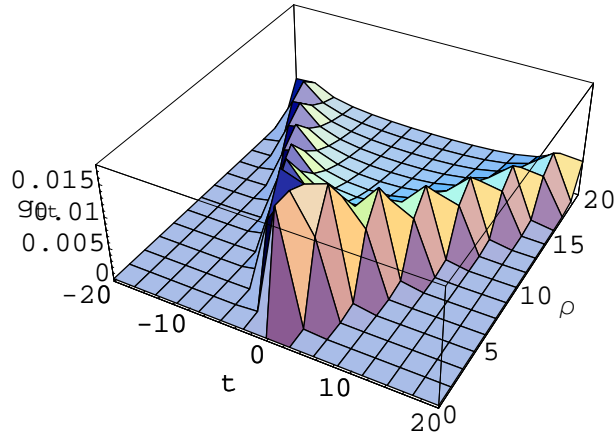}}
\caption{Plot of $g_{tt}$ vs. $\rho$ and $t$ for $D=10$, $p=6$,
$b=1$, $c=1$ and $M_n={1\over 2}$}
\end{figure} \\
\begin{figure}[ht]
\epsfxsize=8cm
\centerline{\epsfbox{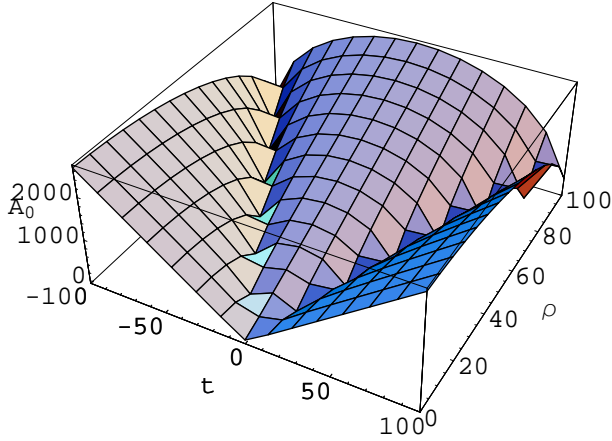}}
\caption{Plot of $A_0$ (where $A_0$=$A_{0,i_1,..,i_p}$) 
vs. $\rho$ and $t$ for same values of the
parameters as mentioned in the previous figure.}
\end{figure} \\
\noindent
Figure ($1$), ($2$), and ($3$) clearly expose the fact that the 
various potentials and the metric components in 
(\ref{metric4})-(\ref{e42}) are smooth for real values of $\rho$ and
$t$. Dilaton profile is also non-singular as can be seen from (\ref{u,Phi}).  
If we take the number of branes $N\rightarrow\infty$, then the
products in (\ref{eu2}) and (\ref{e42}) 
diverge. To get rid of these divergences from the infinite products
we need to multiply divergent constant terms which we call $e^{-2\tilde U_0}$ 
and $e^{-2 K_0}$ respectively. This would allow 
us to remove all the conical singularities along the intermediate line
between two consecutive branes.
$e^{-2\tilde U_0}$ and $e^{-2 K_0}$ can be calculated under the situation
where the potential does not depend on the brane transverse direction
and the mass of each black branes are equal and are equaly spaced
along their common transverse direction. Here we give the values of 
$e^{-2\tilde U_0}$ and $e^{-2 K_0}$ following \cite{Myers} : 
\begin{equation}
e^{-2\tilde U_0}={\prod_{n=1}^\infty}
\Big({{1+{{k}\over{2 n b}}}\over{1-{{k}\over{2 n
b}}}}\Big)^2,
\end{equation}
\begin{equation}
e^{-2 K_0}={\prod_{n=0, m=n+1}^\infty}
\Big({{(n+m+1)^2 b^2}\over {(n+m+1)^2 b^2- k^2}}\Big)^2. 
\end{equation} 
We are therefore able to construct non-singular $Sp$-branes by
analytically continuing certain static black brane solutions. For
particular choices of the dilaton coupling $a$, these solutions can be
embedded in string theory.
\vskip .2cm
\noindent 
In the next part of this note, we construct some new $S0$ brane
configurations starting from non-extremal black dihole solutions.
These non-extremal black dihole solutions are given in \cite{Manko},
\cite{Emparan1}. With the choice of the parameters
(e.g. mass, charge, angular momentum and shifts along the $z$ axis of
each hole) \\
\begin{eqnarray}
m_1&=&m_2 = m,~~~~~~ q_1 = - q_2 = q, \cr
& \cr
a_1&=&a_2 = 0,~~~~~~~~ z_1 = -z_2 = k,
\end{eqnarray}
\noindent
the axisymmetric solution written in cylindrical coordinate, is
\begin{equation}
ds^2 = -e^{2\tilde U} d\tau^2 + e^{-2 \tilde U}[e^{2 K} (d{\rho}^2 +
dz^2)+{\rho}^2 d{\theta}^2],
\label{metric1}
\end{equation}
\begin{eqnarray}
e^{2\tilde U}= {{A^2 - B^2 + C^2}\over{(A + B)^2}},~~~~~~~ e^{2 K} = 
{{A^2 - B^2 + C^2}\over{g r_+ r_- {\tilde r}_+{\tilde r}_-}}, ~~~~~~~~
\Phi = - {{C}\over{A+B}},
\label{gyy,gtt,Phi}
\end{eqnarray}
where $A, B$ and $C$ are given as \\
\begin{eqnarray}
A &=& - (r_+ - r_-)({\tilde r}_+ - {\tilde
  r}_-)[({\kappa^2_+} + {\kappa^2_-})(m^4 + {\kappa_+^2}{\kappa_-^2})- 
4m^2{\kappa_+^2}{\kappa_-^2}] \cr
& \cr 
&+& (r_+ + r_-)({\tilde r}_+ + {\tilde r}_-) ({\kappa_+^2} 
-{\kappa_-^2}) (m^4 + {\kappa_+^2}{\kappa_-^2}) \cr
& \cr
&-&2(r_+ r_- +{\tilde r}_+  {\tilde r}_ )({\kappa_+^2} 
+{\kappa_-^2}) (m^4 - {\kappa_+^2}{\kappa_-^2}), \cr
& \cr
B &=& 4 m {\kappa_+}{\kappa_-}({\kappa_+^2} - {\kappa_-^2})[(r_+ +
r_- + {\tilde r}_+ + {\tilde r}_-){\kappa_+}{\kappa_-} - (r_+ +
r_- - {\tilde r}_+ - {\tilde r}_-) m^2], \cr
& \cr
C &=& 4 k q {\kappa_+}{\kappa_-} [(r_+ - r_- + {\tilde r}_+ 
- {\tilde r}_-)({\kappa_+^2}-m^2){\kappa}_- - (r_+ - r_- - {\tilde r}_+ 
+ {\tilde r}_-)({\kappa_-^2}-m^2){\kappa}_+]. 
\label{ABC}   
\end{eqnarray}
\noindent
The constant $g$ in the expression of $e^{2 K}$ was determined in
\cite{Emparan1} by analysing proper location of the conical
singularity along $z$ axis. 
One can realize this singularity as a strut between the holes and
also outside as a string suspended from infinity. 
$g$, computed in \cite{Emparan1}, is given by 
\begin{equation} 
g= 64\kappa_+^4\kappa_-^4(\kappa_+^2-\kappa_-^2)^2,
\end{equation}
\noindent   
$r_{\pm},   {\tilde r}_{\pm}$ and $\kappa_{\pm}$ are given by
\begin{eqnarray}
r_{\pm} = \sqrt{\rho^2+(z \pm \alpha_+)^2},~~~~~~~~{\tilde r}_{\pm} = 
\sqrt{\rho^2+(z \pm \alpha_-)^2},
\label{R1}
\end{eqnarray}
\begin{eqnarray}
\alpha_{\pm} = \kappa_+ \pm \kappa_-, ~~~~~~~~~~~~~~ {\kappa_{\pm} =
{1\over 2} (\sqrt{m^2 + k^2 + 2k\sqrt{m^2 - q^2}}{\pm}\sqrt{m^2 + k^2 -
  2k\sqrt{m^2 - q^2}})}.
\label{Alpha}
\end{eqnarray}
\noindent
Black holes are lying along the symmetric $z$ axis (where $\rho=0$) centered 
at the points, $z=\kappa_{+}$. Here, $2\kappa_+$ is the distance 
between the center of the two black holes. This distance is
controlled by the parameters $m$, $k$ and $q$. $\kappa_-$ is the
non-extremal parameter in the solutions. 
In the extremal limit $(m=q, \kappa_-=0)$ the 
solutions go back to the standard black dihole solutions. In order to obtain
the extremal solution, one has to go to a suitable coordinate
system \cite{Emparan1}. Metric components diverges when 
$\kappa_-\rightarrow 0$ in the present coordinate system. Black hole 
horizons in this case also lie along the $z$ axis at
$-\kappa_+-\kappa_-\le z \le -\kappa_++\kappa_-$ and 
$\kappa_+-\kappa_-\le z \le \kappa_+ + \kappa_-$. In the extremal
limit, we get degenerate horizons like black dihole as described
in \cite{strominger}. 
$r_{\pm}$, $\tilde r_{\pm}$ in (\ref{R1}) are the radial distances 
to the point of observation from the inner and the outer edge of the 
black holes respectively. \\\\
\noindent
After double Wick rotation ($z\rightarrow{it}$, $\tau\rightarrow{iy}$) 
the solutions (\ref{metric1})-(\ref{R1}) transformed into the
following expressions. \\
\begin{equation}
ds^2 = e^{2\tilde U} dy^2 + e^{-2 \tilde U}[e^{2 K} (d{\rho}^2 -
dt^2)+{\rho}^2 d{\theta}^2],
\label{metric2}
\end{equation}
\begin{eqnarray}
A &=& - (Im~ r)(Re~ \tilde r)[({\kappa^2_+} + {\kappa^2_-})
(m^4 + {\kappa_+^2}{\kappa_-^2}) - 4m^2{\kappa_+^2}{\kappa_-^2}] \cr
& \cr 
&+& (Re~ r)(Re~ \tilde r) ({\kappa_+^2} 
-{\kappa_-^2}) (m^4 + {\kappa_+^2}{\kappa_-^2}) \cr
& \cr
&-&2((Re~ r)^2 + (Im~r)^2 + (Re~ \tilde r)^2 + (Im~\tilde r)^2 )({\kappa_+^2} 
+{\kappa_-^2}) (m^4 - {\kappa_+^2}{\kappa_-^2}), \cr
& \cr
B &=& 4 m {\kappa_+}{\kappa_-}({\kappa_+^2} - {\kappa_-^2})[(Re~r +
Re~\tilde r){\kappa_+}{\kappa_-} - (Re~ r- Re~\tilde r) m^2], \cr
& \cr
C &=& 4 k q {\kappa_+}{\kappa_-} [(Im~r +Im~\tilde r)({\kappa_+^2}
-m^2){\kappa}_- - (Im~r - Im~\tilde r)({\kappa_-^2}-m^2){\kappa}_+].
\label{ABC2}    
\end{eqnarray}
\noindent
Here $\kappa_+$ and $\kappa_-$ in (\ref{Alpha}) certainly do not change 
under these rotations, but $r$ and $\tilde r$ change in the 
following way:
\begin{eqnarray}
r=\sqrt{\rho^2-(t-i\alpha_+)^2},~~~~~~~~~~~\tilde r = \sqrt{\rho^2-
(t-i\alpha_-)^2}.
\label{R2}
\end{eqnarray}
The real nature of our solution can be concluded from the 
following expressions:
\begin{equation}
Re~r =
\sqrt{{1\over2}\Big(\rho^2-t^2+\alpha_+^2 + \sqrt{(\rho^2 -
t^2+\alpha_+^2)^2 + 4t^2 \alpha_+^2}\Big)},  
\end{equation}
\begin{equation}
Im~r = \sqrt{{1\over2}\Big(-\rho^2 - t^2 + \alpha_+^2 +
\sqrt{(\rho^2-t^2+\alpha_+^2)^2 + 4 t^2 \alpha_+^2}\Big)}, 
\end{equation}
\begin{equation}
Re~\tilde r = \sqrt{{1\over2}\Big(\rho^2 - t^2+\alpha_-^2 +
\sqrt{(\rho^2 - t^2 + \alpha_-^2)^2 + 4t^2 \alpha_-^2}\Big)}, 
\end{equation}
\begin{equation}
Im~\tilde r = \sqrt{{1\over2}\Big(-\rho^2 - t^2 + \alpha_-^2 + 
\sqrt{(\rho^2 - t^2 + \alpha_-^2)^2 + 4t^2 \alpha_-^2}\Big)}.
\end{equation}
\noindent
Our solutions can be further generalised by considering static
non-extremal
diholes of four dimensions \cite{Emparan1}, which have non zero dilaton. 
Analytic continuation is similar to what we have done
here. One then gets non-singular $S0$ branes with non trivial
dilaton profile along with electric/magnetic potentials. \\
\vskip .05cm
\noindent
{\bf{Acknowledgments:}} I would like to thank Sudipta Mukherji for
suggesting me this problem and also for his comments on this
manuscript. I would also like to thank Alok Kumar for going through the
manuscript. I also acknowledge many useful discussions with the members 
of our string theory group. \\

\end{document}